# Scan-less hyperspectral dual-comb single-pixel-imaging in both amplitude and phase

Kyuki Shibuya[1, 2], Takeo Minamikawa[2, 3], Yasuhiro Mizutani[2, 4], Hirotsugu Yamamoto[2, 5], Kaoru Minoshima[2, 6], Takeshi Yasui[2, 3, *] and Tetsuo Iwata[2, 3, *]

[1]*Graduate School of Advanced Technology and Science, The Tokushima University, 2-1, Minami-Josanjima, Tokushima 770-8506, Japan*

[2]*JST, ERATO MINOSHIMA Intelligent Optical Synthesizer (IOS) Project, 2-1, Minami-Josanjima, Tokushima, Tokushima 770-8506, Japan*

[3]*Graduate School of Technology, Industrial and Social Sciences, The Tokushima University, 2-1, Minami-Josanjima, Tokushima 770-8506, Japan*

[4]*Graduate School of Engineering, Osaka University, 2-1, Yamadaoka, Suita, Osaka 565-0871, Japan.*

[5]*Center for Optical Research and Education, Utsunomiya University, 7-1-2, Yoto, Utsunomiya, Tochigi 321-858, Japan.*

[6]*Graduate School of Informatics and Engineering, The University of Electro-Communications, 1-5-1, Chofugaoka, Chofu, Tokyo 182-8585, Japan.*

[*]*iwata@tokushima-u.ac.jp, yasui.takeshi@tokushima-u.ac.jp*

-2-# Abstract

We have developed a hyperspectral imaging scheme that involves a combination of dual-comb spectroscopy and Hadamard-transform-based single-pixel imaging. The scheme enables us to obtain 12,000 hyperspectral images of amplitude and phase at a spatial resolution of 46 μm without mechanical scanning. The spectral resolution is 20 MHz, as determined by the linewidth of a single comb mode, and the spectral interval is 100 MHz over a spectral range of 1.2 THz centred at 191.5 THz. As an initial demonstration of our scheme, we obtained spectroscopic images of a standard test chart through an etalon plate. The thickness of an absorptive chromium-coated layer on a float-glass substrate was determined to be 70 nm from the hyperspectral phase images in the near-infrared wavelength region.
-2-

# **Introduction**

An optical frequency comb (OFC) is composed of many optical longitudinal modes that are evenly spaced at a repetition frequency $f_{rep}$ over a wide spectral frequency range. Because all of the individual modes of the OFC can be phase-locked to a frequency standard by precise control of both $f_{rep}$ and a carrier-envelope-offset frequency $f_{ceo}$, the OFC has attracted much attention as an optical frequency ruler [1-3]. Recently, dual-comb spectroscopy (DCS) has emerged as a technique enabling the measurement of mode-resolved spectra of both amplitude and phase in an OFC [4-6]; such measurements are not obtainable using conventional spectrometers. Because of its accurate and rapid measurement capability, many applications of DCS have been developed [7], such as gas analysis [8], strain sensing [9], material characterization [10], polarization measurement [11] spectroscopic ellipsometry [12], and distance measurement [13].

The combination of DCS with an imaging scheme will enable an increase in the number of DCS-based applications. However, few high-speed two-dimensional detectors have a frequency bandwidth of several hundreds of megahertz, which is required to capture the RF interferogram in DCS. Therefore, DCS imaging has been limited to a mechanical scanning scheme that uses a high-speed single-channel photodetector [14]. In this context, development of the DCS imaging without mechanical scanning is desired.

One of the scan-less schemes is to employ single-pixel imaging (SPI). To date, Hadamard transform imaging (HTI) based on the orthogonality of a Hadamard matrix [15] and computational ghost imaging (CGI) based on the statistical randomness of coding mask patterns [16] have been developed. In HTI or CGI, compressive sensing (CS) has been intensively discussed regarding the subject of the sparsity of the structure of the sample object [17]. In SPI, the sample object is coded sequentially by a series of mask patterns generated by a spatial light modulator (SLM) or a digital mirror device (DMD). The corresponding total light intensities transmitted through (or reflected from) the sample object are measured as time-series data by a single-channel photodetector. Next, the original two-dimensional image is reconstructed mathematically by the known coded masks and the measured time-series data. SPI is particularly useful when the appropriate image detector is not readily available. Furthermore, a spatial multiplex advantage is expected to hold like that of the wavelength case in Fourier transform spectroscopy (FTS), provided that the detector noise is dominant and/or the noise is independent of the signal [18, 19]. Previously, CS-based SPI was reported in optical profilometry, where phases of beat signals among inter-modes in the single OFC were used [20]. However, no attempt to combine DCS and SPI has been reported.

In this article, such an attempt is made as a proof of principle: scan-less DCS imaging is reported for the first time, to the best of our knowledge. In this demonstration, an SLM-based HTI for SPI is employed. Pairs of hyperspectral amplitude and phase images of a test chart covered by a Fabry-Pérot etalon plate are measured. From the phase image, the thickness of the absorptive chromium-coated layer on the test chart is determined.

## Result

### *Experimental procedure*

A schematic of the proposed DCS-SPI system is shown in Fig. 1 to illustrate how the Hadamard transform (HT)-based SPI scheme is incorporated into the typical DCS scheme. In the DCS system, two laser systems, each comprising a mode-locked erbium-doped fibre-laser oscillator and an erbium-doped amplifier in the 1.56-μm band, were used as a signal OFC ($f_{ceo1}$ = 10.5 MHz, $f_{rep1}$ ~ 100 MHz) and a local OFC ($f_{ceo2}$ = 10.5 MHz, $f_{rep2}$ ~ 99,999,525 Hz, $\Delta f_{rep}$ = $f_{rep1}$ - $f_{rep2}$ ~ 475 Hz). These laser systems were tightly and coherently locked to each other with a certain frequency offset $\Delta f_{rep}$ using a narrow-linewidth, external-cavity, external-cavity laser diode (ECLD) as an intermediate laser [10, 21]; this coherent locking allows us to perform coherent averaging (accumulation) of the interferograms observed in DCS [22, 23].

The procedure used to perform the measurement is as follows. The signal-OFC light beam passing through the sample object is superimposed and interfered with the local OFC light through the use of polarisers (PBS1 and PBS2) and a half-wave plate (HWP). An optical bandpass filter (BPF) is used to narrow the spectral bandwidth of the two OFC light beams to prevent aliasing in DCS. Next, the interfered light is directed to a reflection-type liquid-crystal-on-silicon SLM (LCOS-SLM) device placed at the image plane of the sample object. The intensity distribution of the sample image is spatially modulated as a black-and-white binary pattern by a combination of the LCOS-SLM with PBS2. Finally, the spatially modulated interferogram is acquired as the time-series data by a single-channel photodetector (frequency bandwidth = 100 MHz) and a 14-bit digitizer (sampling rate = $10^8$ samples/s).

Figure 2 shows a series of reconstruction procedures in DCS-SPI, for which a cyclic HTI scheme [24] was used. From every row of an $N \times N$-sized Hadamard matrix $\mathbf{M}(x,y)$, $N$ sheets of $n \times n$-sized coding masks were generated, where $n^2 = N$. Next, a set of $N$ interferograms $I_k(t)$ ($k$; 1-$N$) were measured for the individual coding masks. Discrete Fourier transforms (DFTs) of $I_k(t)$ were performed to provide the spectra $S_k(f)$ as a function of a frequency $f$, whose real parts are $S_{k\_r}(f)$ and imaginary parts are $S_{k\_i}(f)$. As a result, $N$ sets of real and imaginary frequency spectra were obtained, corresponding to the $N$ coding masks. For the pairs of $N$ spectra, we produced the column vectors $\mathbf{S}_r(f_q)$ and $\mathbf{S}_i(f_q)$, ($q$; 1-$m$). Subsequently, we calculated and derived pairs of the real and the imaginary column vectors $\mathbf{O}_r(f_q)$ and $\mathbf{O}_i(f_q)$ corresponding to the frequency $f_q$ using the inverse Hadamard transform (IHT). From the two $N \times 1$-sized column vectors, a pair of real and imaginary images, $O_r(f_q; x,y)$ and $O_i(f_q; x,y)$, were derived. Finally, $m$ sets of amplitude and phase images were obtained [25].

To demonstrate the basic performance of the DCS-SPI for acquiring hyperspectral amplitude and phase images, we used a 1951-USAF test chart (#38-256, Edmund Optics Inc., USA) consisting of a 1.5-mm-thick piece of float glass on which a chromium layer was vacuum-deposited to form various negative line patterns. To provide the test chart with a deliberate spectroscopic variation in transmittance, we superimposed a Fabry-Pérot etalon plate (Koshin Kogaku Co., Ltd., Japan; centre transmission wavelength and free spectral range (FSR) of 1,550 nm and 90 GHz, respectively) onto the test chart.

***Mode-resolved spectroscopy of both amplitude and phase***

Figure 3(a) shows an interferogram pulse train obtained without the Hadamard mask, where the number of accumulations is 250 and the data acquisition time is 2.63 sec. The abscissa axis is scaled by both the laboratory time and the effective time via the time-scale-expansion factor of $f_{rep1}/\Delta f_{rep} \sim 210,526$. In the time window of 50 ns, we observe five consecutive interferograms temporally spaced by $1/f_{rep1}$. Figure 3(b) shows the same temporal interferogram, but magnified. We observe multiple decaying echo pulses at a period of 11 ps, as determined by 1/FSR of the Fabry-Pérot etalon. The decay time of the envelope is ~80 ps, which is determined by the inverse of the full width at half-maximum (FWHM) of the entire transmission band of the etalon. Figure 3(c) shows a further magnified interferogram shown in Fig. 3(b). We observe temporal behaviour of the interferogram with high SNR because of the tight locking of the pair of OFCs [10, 21] and the coherent averaging [22, 23].

Next, we derived a mode-resolved spectrum [Fig. 3(d)] from the interferogram shown in Fig. 3(a) via Fourier transform, resulting in the amplitude spectrum (blue line) and phase spectrum (red line) shown in the figure. In the amplitude spectrum, we observe thirteen resonance transmission peaks within the spectral range from 191.0 to 192.2 THz that are determined by the bandwidth of the BPF inserted to prevent aliasing in the DCS. The frequency interval is again 90 GHz and is identical to the FSR of the etalon. In each spectral band transmitted, we observe multiple comb modes, for example, from 191.5040 to 191.5050 THz, as shown in Fig. 3(e). Although the amplitude is expected to be zero at the frequency gap region among the individual comb modes, a slightly non-zero signal appears. This signal may originate from the residual timing jitter between the two OFCs [26]. The phase spectrum shown in Figs. 3(d) or 3(e) exhibits the chirp between the signal-OFC light and the local-OFC light. The FSR and the finesse at 191.6 THz were determined as 89.98 GHz and 7.05, respectively, which agreed well with the manufacturer's specifications; i.e., our DCS system was found to operate satisfactorily.

*Mode-resolved amplitude imaging*

Next, we attempted to derive a hyperspectral amplitude image by incorporating the HT-based SPI technique. Figure 4(a) shows the photograph of the test chart, in which the red square box represents the region of interest (ROI; 1.47 mm × 1.47 mm) for reconstructing the image. In the ROI (Group 1, Element 6 in the 1951-USAF test chart), three horizontal lines (each of which has a width of 0.14 mm) transmit light due to its negative pattern. The area of contact between the etalon and the test chart is depicted in light green. DCS-HTI was conducted with $N = 1,024$ and $n = 32$. The size of the unit pixel of the coding mask was 46 μm × 46 μm.

Figures 4(b)-i to vi show the reconstructed images at $f_1 = 191.14034$, $f_2 = 191.14040$, $f_3 = 191.59286$, $f_4 = 191.59292$, $f_5 = 191.63260$, and $f_6 = 191.63266$ THz, respectively. As shown in Fig. 4(c), which is the same as Fig. 3(d) but expanded in part for the following explanations, the frequencies $f_1$ and $f_2$ are located at the tail of the lower frequency side of the transmission band of the etalon. $f_1$ is positioned at the peak of the comb mode, and $f_2$ is positioned at the valley. The frequencies $f_3$ and $f_4$ are located at the centre frequency of the transmission band of the etalon. $f_3$ and $f_4$ are positioned at the peak and valley, respectively, of the comb mode. The frequencies $f_5$ and $f_6$ are located at the tail of the higher frequency side of the transmission band of the etalon. $f_5$ is positioned at the peak, and $f_6$ is positioned at the valley. Each amplitude image has been background-subtracted to set the intensity of the chromium-coated part of the image to zero; the image obtained from the chromium-coated area of the same test chart was measured in advance. The reason for this subtraction procedure is that the near-infrared light slightly transmits through the thin chromium layer, resulting in a dc bias in the reconstructed image. The ordinate

scale of each reconstructed image was normalized by the maximum value in the image.

The $f_3$ image shown in Fig. 4(b)-iii is fairly well spatially resolved, whereas the signal-to-noise ratios of the $f_1$ and $f_5$ images shown in Figs. 4(b)-i and 4(b)-v, respectively, are degraded because of the lower illumination intensity. The $f_2$ and $f_6$ images shown in Figs. 4(b)-ii and 4(b)-vi, respectively, contain no information, as expected. However, the image shown in Fig. 4(b)-iv exhibits some line structure despite $f_4$ being at the mode gap; this observation is due to the cross-talk. The results indicate that we successfully observed a spectroscopic amplitude image. The spatial resolution and the spectral resolution are 46 μm and 20 MHz, respectively, as determined by the unit pixel size of coding mask and the linewidth of the OFC mode, respectively. Although we show only six images here, 12,000 images over the spectral range from 191.0 to 192.2 THz were obtained.

*Mode-resolved phase imaging*

Next, we obtained phase images of the test chart at the six optical frequencies. Figures 5-i to vi show the $f_1 \sim f_6$ phase images, which correspond to the $f_1 \sim f_6$ amplitude images already shown in Figs. 4(b)-i to vi. Here, we compensated the phase image by the "null" phase image that was obtainable without the test chart; i.e., the real and imaginary parts of the null image were subtracted. The chromium-background subtraction procedure conducted for the amplitude images degraded the signal-to-noise ratio of these phase images because of the low transmittance of the chromium layer. We can observe the line structure for the $f_1$, $f_3$, and $f_5$ phase images, although the phase contrast is relatively low. For the $f_2$, $f_4$, and $f_6$ phase images, no structure appears because of the low illumination light intensity and the resulting random phase noise. The difference in the phase value in the $f_3$ image is caused by the optical thickness of the test chart, corresponding to the difference of the optical thickness between the chromium-coated and the uncoated part of the test chart. Although the transmittance of the chromium-coated layer is relatively low, the weak transmitted light allows the phase image to be obtained. If the transmittance were exactly zero, then the phase values would be randomly distributed from -π to π. This experimental result reveals the sensitivity and the power of the phase image measurement in DCS-SPI.

Figures 6(a) and 6(b) show cross-sectional profiles of the $f_3$ amplitude and phase images along the red line A-A' drawn in Figs. 4(b)-iii and 5-iii, respectively. The chromium layer is shown in grey. The profile of the test chart is clearly reconstructed by the amplitude image, although the border between the coated and uncoated part is blurred because of the insufficient spatial resolution (46 μm) of SPI. The phase profile shows nearly the same behaviour as the amplitude profile. The difference in phase represents the time for light to propagate through the chromium layer. Taking into account the refractive index of the chromium-coated layer [27], which is 3.66 at 1.61 μm, the optical thickness of the layer was estimated to be approximately 70 nm. To verify this result, we measured the surface profile of the same test chart using an atomic force microscope (OLS3500-PTU, Olympus Corp., Japan, depth resolution; 1.0 nm) and confirmed that the height of the coated layer was 70.0 nm. The phase difference corresponding to the 70.0-nm-thick layer at $f_3$ is depicted by two parallel dotted lines in Fig. 6(b). The difference in the phase value is consistent with that obtained from the phase measurement in DCS-SPI.

## Discussions

The features of the proposed scheme of DCS-SPI enable us to obtain hyperspectral phase images in addition to conventional hyperspectral amplitude images, which opens the door to super-fine hyperspectral imaging. The use of the phase information obtained from DCS can effectively obtain the optical constants and thicknesses of transparent or semi-transparent materials. Furthermore, the use of both amplitude and phase information enables the hyperspectral imaging of complex refractive index or complex permittivity without the need for the Kramers-Kronig relation. The scan-less scheme, which avoids mechanical moving parts, enables more precise measurements. Furthermore, the use of a single-channel detector brings more versatility in the choice of the detector in terms of the spectroscopic sensitivity, the response time, and the cost. The minimum time required for obtaining a single interferogram is determined by $1/\Delta f_{rep}$ and that required for reconstructing the final image is determined by $N/\Delta f_{rep}$.

DCS-SPI and the conventional pixel-scanning (PS)-DCS imaging approach [14] appear to have no significant difference in principle from the viewpoint of the acquisition time of the image. However, the acquisition time of PS-DCS imaging is strongly dependent on the scan mechanism employed. One feature of DCS-SPI is that it provides a spatial multiplex advantage as well as the conventional FTS in the wavelength, provided that the detector noise is dominant. The spatially averaged, signal-detection scheme in DCS-SPI might be effective for eliminating the spurious noise that occurs at a specified pixel of the image; i.e., the multiplex advantage is maintained when the noise is independent of the signal. The acquisition time could be shortened markedly if the CS imaging technique is introduced into DCS-SPI for observing a spatially sparse sample object.

In summary, we demonstrated scan-less hyperspectral imaging in amplitude and phase by combining DCS with cyclic HT-based SPI at a spatial resolution of 46 μm and a spectral resolution of 20 MHz (determined by the linewidth of a single OFC mode). Using the proposed system, 12,000 spectral images in pairs of amplitude and phase were obtained with a spectral interval of $f_{rep1}$ (= 100 MHz) over a spectral range of 1.2 THz centred at 191.5 THz. Notably, the thickness of a 70-nm coated chromium layer on the sample object was determined from the reconstructed phase images. We believe that this demonstration is the first application of DCS-SPI in terms of phase. The maximum thickness to be measured is limited by the OFC wavelength (~1.5 μm) because of phase wrapping ambiguity. However, hyperspectral phase images are expected to further expand the dynamic range by introducing a synthetic wavelength in the OFC modes. Although the proof of principle demonstration was conducted in the near-infrared wavelength region, the approach is expected to be applicable in other spectral regions, such as the ultraviolet [28], the visible [29], the mid-infrared [5, 22], and the THz regions [6, 30] because of the versatility of DCS-SPI.

## Methods

*Experimental configuration*

A schematic of the proposed DCS-SPI system is shown in Fig. 1, where the HT-based SPI scheme is incorporated into the typical DCS scheme. Two mode-locked erbium-doped fibre-laser oscillator and amplifier systems (OCLS-HSC-D100-TKSM, NEOARK CORP., Japan) were used as signal and local OFCs. The repetition frequency of the signal comb was $f_{rep1}$ ~ 100 MHz, and that of the local comb was $f_{rep2}$ ~ 99.999,525 MHz, i.e., the difference between the two frequencies was $\Delta f_{rep}$ ~ 475 Hz. The centre wavelengths, the spectral bandwidths, and the average power of the laser systems were 1,560 nm, 20 nm, and 120 mW, respectively. The carrier-envelope offset frequency $f_{ceo1}$ and the $f_{rep1}$ of the signal comb were phase-locked to a rubidium frequency standard (FS725, Stanford Research Systems, USA, accuracy: $5 \times 10^{-11}$; instability: $2 \times 10^{-11}$ at 1 sec) by controlling the drive current of the pumping laser and the cavity length of the fibre laser, respectively. To achieve tight and coherent locking of the local OFC to the signal OFC, a narrow-linewidth, external-cavity, CW laser diode (ECLD, PLANEX, Redfern Integrated Optics Inc., USA, centre wavelength: 1,550 nm; FWHM: <2.0 kHz) was phase-locked to one of the single modes in the local OFC. In addition, one of the single modes in the signal OFC was phase-locked to the same ECLD by controlling $f_{rep1}$ by an electro-optics modulator (Model 4004, New Focus, Newport Corp., USA, frequency bandwidth: ~100 MHz) while maintaining the phase-lock of $f_{ceo1}$ to the frequency standard [10, 21]. In this configuration, the relative linewidth between the signal and the local OFC could be narrowed to less than 1 Hz [21], which allowed us to perform coherent averaging (accumulation) of the interferograms observed in DCS [22, 23].

The horizontally polarized signal-OFC light beam passing through a sample object was superimposed with the vertically polarized local OFC light beam via a polarization beam splitter (PBS1). An optical BPF (FB1570-12, Thorlabs, Inc., USA, centre wavelength: 1570 nm; bandwidth: 12 nm) was used to narrow the spectral bandwidth of the two OFC light beams to prevent aliasing in DCS. An HWP (WPH05M-1550, Thorlabs, Inc., USA, design wavelength: 1550 nm; zero-order) was used to rotate their polarization directions by 45 degrees. The second polarization beam splitter (PBS2) transmits the horizontally polarized components of the superimposed light, through which the two OFC light beams are interfered. Next, the interfered light was incident on a reflection-type LCOS-SLM (SLM100-04, Santec Corp., Japan, number of pixels: 1440×1050; pixel size: 10.0 μm × 10.0 μm; response time: 100 ms). The magnification factor of the image on the SLM relative to the sample object was 2.5, which was determined by focal lengths of a pair of lenses L1 and L2 (L1, focal length: $F_1$ = 100 mm; L2, focal length: $F_2$ = 250 mm). As shown by a black-and-white pattern on LCOS-SLM, a series of binary-phased masks consisting of 0 and π modulated the incident light beams, generating the vertically polarized light components modulated in the same manner as the binary-phased masks, i.e., the intensity distribution of the sample image was spatially modulated. The vertically polarized light component was reflected by PBS2 at this time and then passed through a ×1/3 beam expander composed of lenses L3 and L4 (L3, focal length: $F_3$ = 150 mm; L4, focal length: $F_4$ = 50 mm) and a ×10 objective lens (NA: 0.25). Finally, the spatially modulated interferogram was detected by an InGaAs PIN photodiode (PDB415C, Thorlabs, Inc., USA,, spectroscopic sensitivity: 800-1,700 nm; frequency bandwidth: 100 MHz) as the time-series data and fed into a 14-bit digitizer (NI PXIe-5122, National Instruments Corp., USA, sampling rate: 99,999,525 samples/s; number of sampling points: 1,052,626).

### *Cyclic HTI*

We used a cyclic HTI scheme [24] for DCS-SPI. Figure 2 shows a series of reconstruction procedures in DCS-SPI. From every row of an $N×N$-sized Hadamard matrix **M**(*x*,*y*), $N$ sheets of $n×n$-sized coding masks were generated, where $n^2 = N$. Next, a set of $N$ interferograms $I_k(t)$ ($k$; 1-$N$) were measured for the individual coding masks. Strictly speaking, as described in ref. [31], the first row and the first column of the matrix $\mathbf{M}(x,y)$ become zero vectors when making the matrix cyclic: the first coding mask becomes a uniformly dark pattern that has no meaning, and the (1,1)-pixel elements of the individual reconstructed images contain no information. DFTs of $I_k(t)$ give their spectra $S_k(f)$ as a function of a frequency $f$, whose real parts are $S_{k\_r}(f)$ and imaginary parts $S_{k\_i}(f)$. As a result, $N$ sets of real and imaginary frequency spectra were obtained corresponding to the $N$ coding masks. For the pairs of $N$ spectra, we constructed column vectors, $\mathbf{S}_r(f_q)$ and $\mathbf{S}_i(f_q)$, ($q$; 1-$m$). Next, we calculated and derived pairs of the real and the imaginary column vector corresponding to the frequency $f_q$ by IHT: $\mathbf{O}_r(f_q) = (2/N)\mathbf{M}^T\mathbf{S}_r(f_q) = 2\mathbf{M}^{-1}\mathbf{S}_r(f_q)$, and $\mathbf{O}_i(f_m) = (2/N)\mathbf{M}^T\mathbf{S}_i(f_q) = 2\mathbf{M}^{-1}\mathbf{S}_i(f_q)$, where $\mathbf{M}^T$ is the transpose of the cyclic Hadamard matrix $\mathbf{M}$ and is proportional to its inverse: $\mathbf{M}^T/N = \mathbf{M}^{-1}$. From the two $N\times 1$-sized column vectors, $\mathbf{O}_r(f_q)$ and $\mathbf{O}_i(f_q)$ corresponding to the frequency $f_q$, a pair of real and imaginary images, $O_r(f_q; x,y)$ and $O_i(f_q; x,y)$, respectively, were derived. Next, we calculated and derived pairs of the real and the imaginary column vector $\mathbf{O}_r(f_q)$ and $\mathbf{O}_i(f_q)$, respectively, corresponding to the frequency $f_q$ by IHT. From the two $N\times 1$-sized column vectors, a pair of real and imaginary images, $O_r(f_q; x,y)$ and $O_i(f_q; x,y)$, respectively, were derived. Finally, $m$ sets of amplitude and phase images were obtained as follows [25]:

$$O_{\mathrm{amp.}}\left(f_q;\ x,y\right) = \sqrt{O_r\left(f_q;\ x,y\right)^2 + O_i\left(f_q;\ x,y\right)^2} \quad , \tag{1}$$

$$O_{\mathrm{phase}}\left(f_q;\ x,y\right) = \tan^{-1}\left[O_i\left(f_q;\ x,y\right)/O_r\left(f_q;\ x,y\right)\right]. \tag{2}$$

## Acknowledgements

This work was supported by grants for the Exploratory Research for Advanced Technology (ERATO) MINOSHIMA Intelligent Optical Synthesizer (IOS) Project (JPMJER1304) from the Japan Science and Technology Agency and a Grant-in-Aid for JSPS Research Fellow No. 16J08197 from the Ministry of Education, Culture, Sports, Science, and Technology of Japan. The authors acknowledge Dr. Yoshiaki Nakajima and Dr. Akifumi Asahara of The University of Electro-Communications, Japan for their help in constructing the dual-comb sources. The authors also acknowledge Ms. Shoko Lewis of Tokushima Univ., Japan for her help in the preparation of the manuscript.

## Author contributions

T. I., T. Y., and K. M. conceived the project. K. S. performed the experiments and analysed the data. K. S., T. Y., and T. I wrote the manuscript. T. M., Y. M., and H. Y. discussed the results and commented on the manuscript.

## Competing financial interests statement

The authors declare no competing financial interests.

# Figure legends

Fig. 1. Schematic diagram of the DCS-SPI system. ECLD is a narrow-linewidth external-cavity laser diode; M is a mirror; PBS1 and PBS2 are polarization beam splitters; L1, L2, L3, and L4 are lenses; BPF is an optical bandpass filter; HWP is a half-wave plate; and LCOS-SLM is a reflection-type liquid-crystal-on-silicon device.

Fig. 2. Data processing for obtaining $m$ pairs of mode-resolved amplitude and phase images.

Fig. 3. Interferogram and mode-resolved spectrum obtained from a DCS measurement. The sample object was a standard test chart on which an etalon plate was superimposed. (a) An interferogram pulse train measured without a Hadamard mask; (b) the same interferogram as (a) but expanded in time; (c) the same interferogram further expanded; (d) a mode-resolved amplitude spectrum (blue line) and a phase spectrum (red line) derived from the interferogram (a) by Fourier transform; and (e) the same as (d) but expanded in frequency.

Fig. 4. (a) Photograph of a test chart; the red-square box represents the ROI, and the area on which an etalon was superimposed is depicted in light green. (b) Mode-resolved amplitude images: (i) $f_1$ = 191.14034, (ii) $f_2$ = 191.14040, (iii) $f_3$ = 191.59286, (iv) $f_4$ = 191.59292, (v) $f_5$ = 191.63260, and (vi) $f_6$ = 191.63266 THz. (c) Re-plot of the amplitude spectrum shown in Fig. 3(d).

Fig. 5. Mode-resolved phase images: (i) $f_1$ = 191.14034, (ii) $f_2$ = 191.14040, (iii) $f_3$ = 191.59286, (iv) $f_4$ = 191.59292, (v) $f_5$ = 191.63260, and (vi) $f_6$ = 191.63266 THz.

Fig. 6. Cross-section profiles of the line A-A' drawn on the reconstructed images shown in Fig. 4(b)-iii

and Fig. 5-iii at $f_3$ = 191.59286 THz: (a) amplitude and (b) phase.

# References

[1] Udem, Th., Reichert, J., Holzwarth, R. & Hänsch, T. W. Accurate measurement of large optical frequency differences with a mode-locked laser. *Opt. Lett.* **24**, 881-883 (1999).

[2] Niering, M. *et al.* Measurement of the hydrogen 1S-2S transition frequency by phase coherent comparison with a microwave cesium fountain clock. *Phys. Rev. Lett.* **84**, 5496-5499 (2000).

[3] Udem, Th., Holzwarth, R. & Hänsch, T. W. Optical frequency metrology. *Nature* **416**, 233-237 (2002).

[4] Schiller, S. Spectrometry with frequency combs. *Opt. Lett.* **27**, 766–768 (2002).

[5] Keilmann, F., Gohle, C. & Holzwarth, R. Time-domain mid-infrared frequency-comb spectrometer. *Opt. Lett.* **29**, 1542-1544 (2004).

[6] Yasui, T. *et al.* Terahertz frequency comb by multifrequency-heterodyning photoconductive detection for high-accuracy, high-resolution terahertz spectroscopy. *Appl. Phys. Lett.* **88**, 241104 (2006).

[7] Coddington, I., Newbury, N. & Swann, W. Dual-comb spectroscopy. *Optica* **3**, 414-426 (2016).

[8] Coddington, I., Swann, W. C. & Newbury, N. R. Time-domain spectroscopy of molecular free-induction decay in the infrared. *Opt. Lett.* **35**, 1395-1397 (2010).

[9] Kuse, N., Ozawa, A. & Kobayashi, Y. Static FBG strain sensor with high resolution and large dynamic range by dual-comb spectroscopy. *Opt. Express* **21**, 11141-11149 (2013).

[10] Asahara, A. *et al.* Dual-comb spectroscopy for rapid characterization of complex optical properties of solids. *Opt. Lett.* **41**, 4971-4974 (2016).

[11] Sumihara, S. *et al.* Polarization-sensitive dual-comb spectroscopy. J. Opt. Soc. Am. B **34**, 154-159 (2017).

[12] Minamikawa, T. *et al.* Dual-optical-comb spectroscopic ellipsometry. *Conference on Lasers and Electro-Optics 2016*, SW1H.5 (2016).

[13] Coddington, I., Swann, W. C., Nenadovic, L. & Newbury, N. R. Rapid and precise absolute distance

measurements at long range, *Nature Photon.* **3**, 351-356 (2009).

[14] Ideguchi, T. *et al.* Coherent Raman spectro-imaging with laser frequency combs. *Nature* **502**, 355-358 (2013).

[15] Pratt, W. K., Kane, J. & Andrews, H. C. Hadamard transform image coding. *Proc. IEEE* **57**, 58 (1969).

[16] Shapiro, J. H. Computational ghost imaging. *Phys. Rev. A* **78**, 061802 (2008).

[17] Duarte, M. F. *et al.* Single-pixel imaging via compressive sampling. *IEEE Signal Process. Mag.* **25**, 83-91 (2008).

[18] Clemente, P. *et al.* Compressive holography with a single-pixel detector. *Opt. Lett.* **38**, 2524-2527 (2013).

[19] Shibuya, K., Nakae, K., Mizutani, Y. & Iwata, T. Comparison of reconstructed images between ghost imaging and Hadamard transform imaging. *Opt. Rev.* **22**, 897-902 (2015).

[20] Pham, Q. D. & Hayasaki, Y. Optical frequency comb interference profilometry using compressive sensing. *Opt. Express* **21**, 19003-19011 (2013).

[21] Nishiyama, A. *et al.* Doppler-free dual-comb spectroscopy of Rb using optical-optical double resonance technique. *Opt. Express* **24**, 25894-25904 (2016).

[22] Baumann, E. *et al.* Spectroscopy of the methane $\nu_3$ band with an accurate midinfrared coherent dual-comb spectrometer. *Phys. Rev. A* **84**, 062513 (2011).

[23] Roy, J. Deschênes, J.-D. Potvin, S. & Genest, J. Continuous real-time correction and averaging for frequency comb interferometry. *Opt. Express* **20**, 21932-21939 (2012).

[24] Harwit, M. & Sloane, N.J.A. *Hadamard Transform Optics* (Academic Press Inc., Ltd., New York, 1979).

[25] Mizuno, T. & Iwata, T. Hadamard-transform fluorescence-lifetime imaging. *Opt. Express* **24**, 8202-8213 (2016).

[26] Yasui, T. *et al*. Adaptive sampling dual terahertz comb spectroscopy using dual free-running femtosecond lasers, *Sci. Rep.* **5**, 10786 (2015).

[27] Johnson, P. B. & Christy, R. W. Optical constants of transition metals: Ti, V, Cr, Mn, Fe, Co, Ni, and Pd. *Phys. Rev. B* **9**, 5056-5070 (1974).

[28] Bernhardt, B. *et al.* Cavity-enhanced dual-comb spectroscopy. *Nature Photon.* **4**, 55-57 (2010).

[29] Ideguchi, T., Poisson, A., Guelachvili, G., Hänsch, T. W. & Picqué, N. Adaptive dual-comb spectroscopy in the green region. *Opt. lett*. **37**, 4847-4849 (2012).

[30] Hsieh, Y. D. *et al.* Terahertz comb spectroscopy traceable to microwave frequency standard. *IEEE Trans. on THz Sci. Tech.* **3**, 322-330 (2013).

[31] Tetsuno, S., Shibuya, K. & Iwata, T. Subpixel-shift cyclic-Hadamard microscopic imaging using a pseudo-inverse-matrix procedure. *Opt. Express* **25**, 3420-3432 (2017).

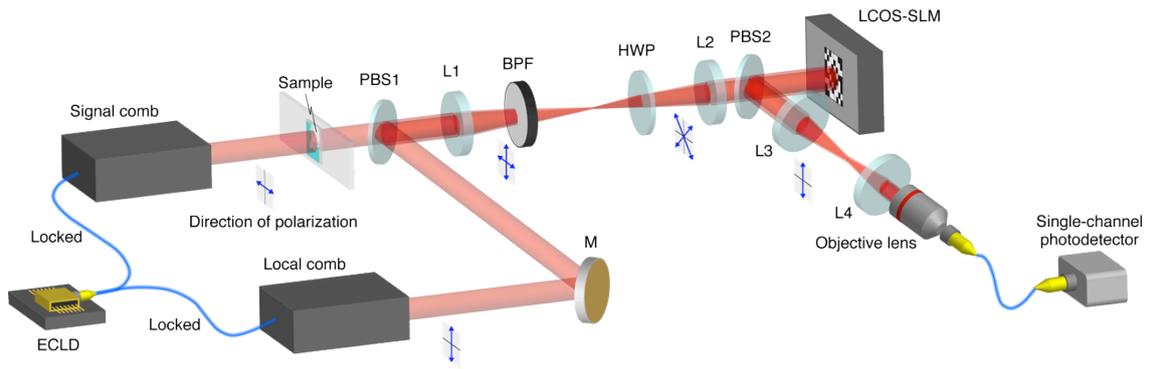

Fig. 1. Schematic diagram of the DCS-SPI system. ECLD is a narrow-linewidth external-cavity laser diode; M is a mirror; PBS1 and PBS2 are polarization beam splitters; L1, L2, L3, and L4 are lenses; BPF is an optical bandpass filter; HWP is a half-wave plate; and LCOS-SLM is a reflection-type liquid-crystal-on-silicon device.

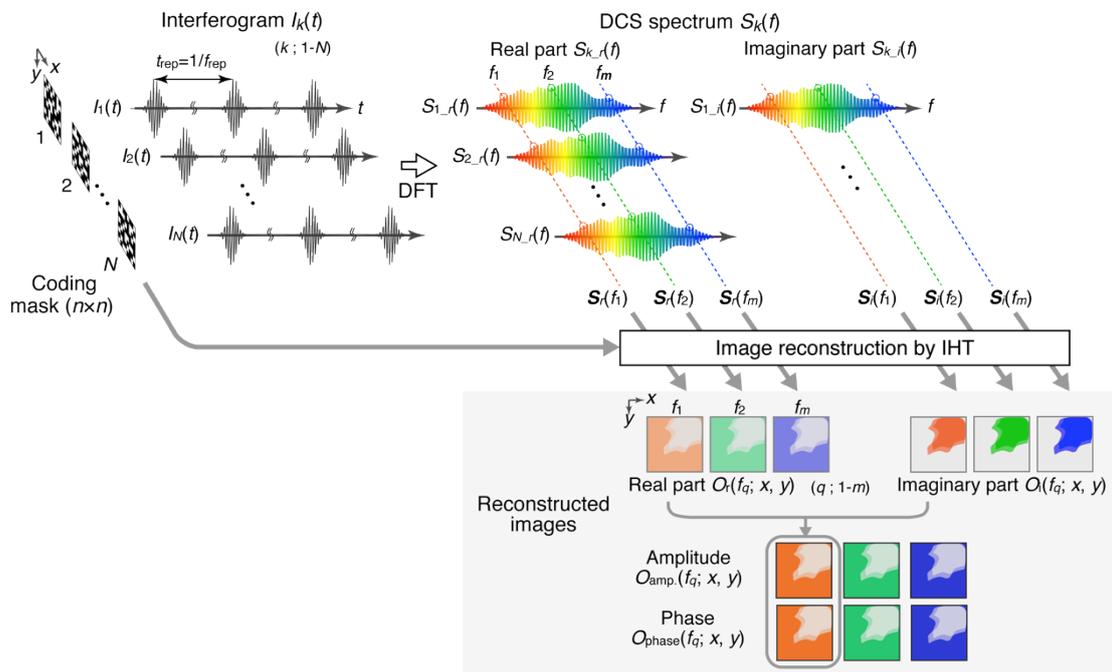

Fig. 2. Data processing for obtaining *m* pairs of mode-resolved amplitude and phase images.

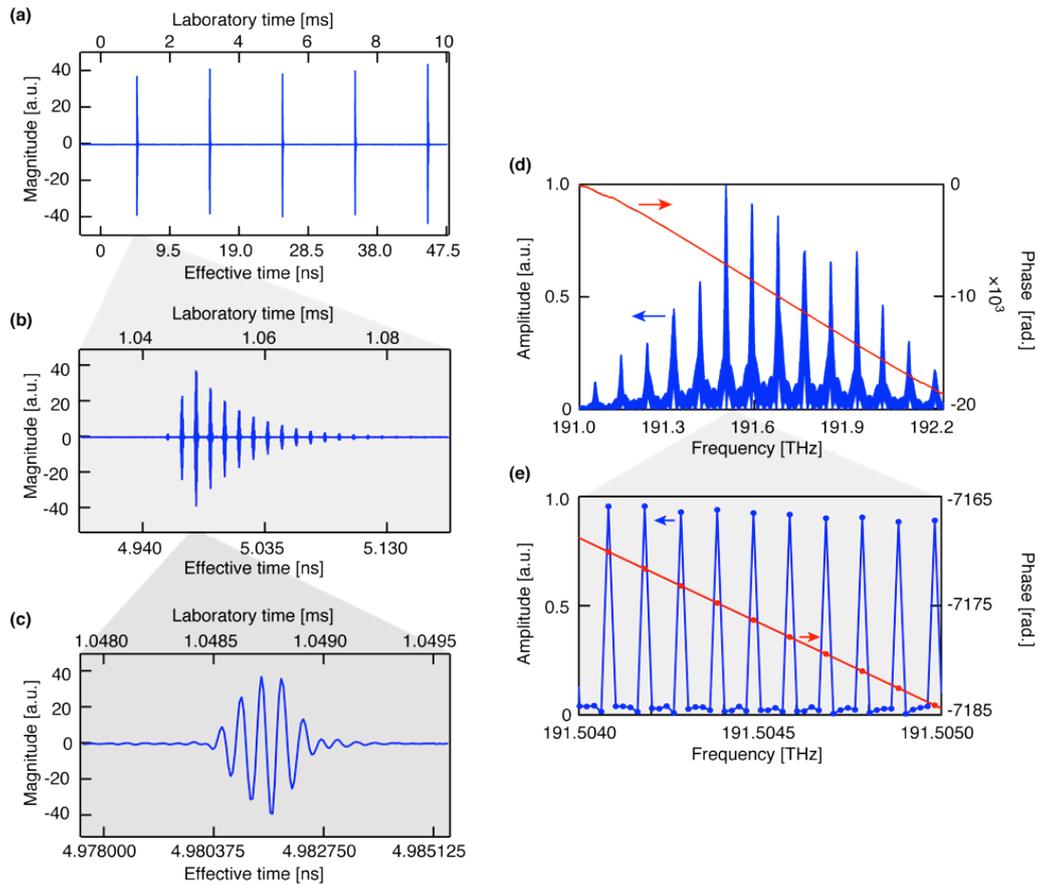

Fig. 3. Interferogram and mode-resolved spectrum obtained from a DCS measurement. The sample object was a standard test chart on which an etalon plate was superimposed. (a) An interferogram pulse train measured without a Hadamard mask; (b) the same interferogram as (a) but expanded in time; (c) the same interferogram further expanded; (d) a mode-resolved amplitude spectrum (blue line) and a phase spectrum (red line) derived from the interferogram (a) by Fourier transform; and (e) the same as (d) but expanded in frequency.

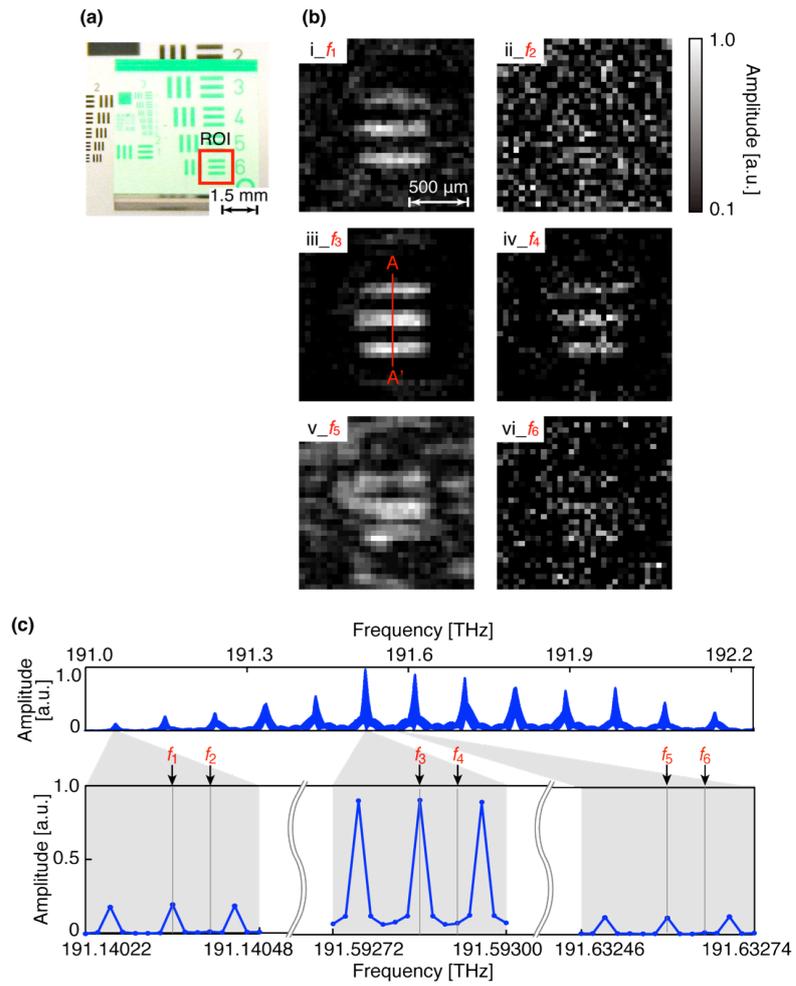

Fig. 4. (a) Photograph of a test chart; the red-square box represents the ROI, and the area on which an etalon was superimposed is depicted in light green. (b) Mode-resolved amplitude images: (i) $f_1$ = 191.14034, (ii) $f_2$ = 191.14040, (iii) $f_3$ = 191.59286, (iv) $f_4$ = 191.59292, (v) $f_5$ = 191.63260, and (vi) $f_6$ = 191.63266 THz. (c) Re-plot of the amplitude spectrum shown in Fig. 3(d).

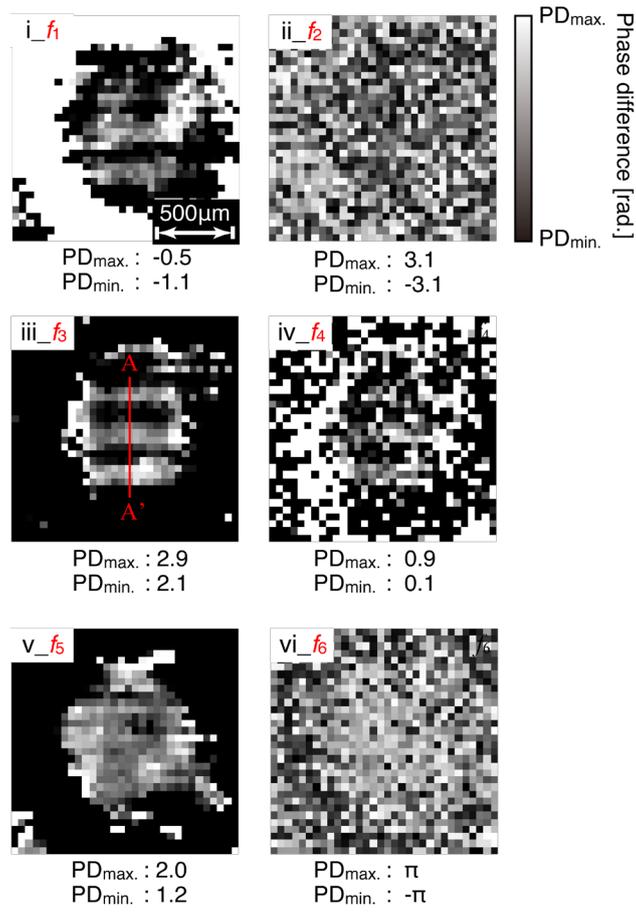

Fig. 5. Mode-resolved phase images: (i) $f_1$ = 191.14034, (ii) $f_2$ = 191.14040, (iii) $f_3$ = 191.59286, (iv) $f_4$ = 191.59292, (v) $f_5$ = 191.63260, and (vi) $f_6$ = 191.63266 THz.

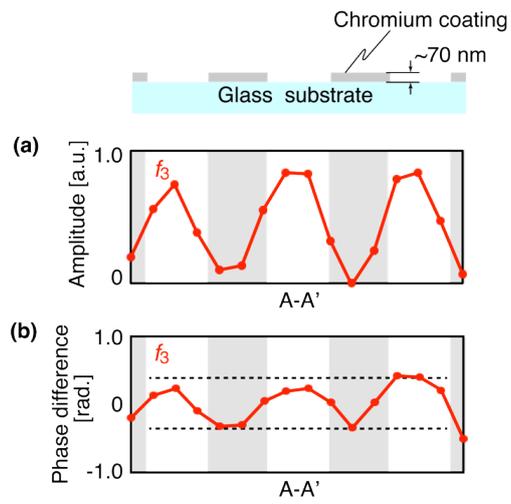

Fig. 6. Cross-sectional profiles of the line A-A' drawn on the reconstructed images shown in Fig. 4(b)-iii and Fig. 5-iii at $f_3$ = 191.59286 THz: (a) amplitude and (b) phase.